\documentclass{article}
\usepackage{spconf,amsmath,amssymb,graphicx}
\usepackage{gensymb}
\usepackage{glossaries}
\usepackage{booktabs}
\usepackage{multirow}
\usepackage{bm} 
\usepackage{subfloat}
\usepackage{subcaption}
\usepackage{flushend}
\usepackage[maxbibnames=4,doi=false,style=ieee,isbn=false,url=false,eprint=false]{biblatex}
\usepackage{enumitem}
\usepackage{balance}

\newcommand{\HR}{\mathrm{\mathbf{I}_{HR}}}
\newcommand{\LR}{\mathrm{\mathbf{I}_{LR}}}
\newcommand{\HRt}{\mathrm{\mathbf{T}_{HR}}}
\newcommand{\THR}{\mathcal{T}(\mathrm{\mathbf{I}_{HR}})}

\newcommand{\TLR}{\mathcal{T}(\mathrm{\mathbf{I}_{LR}})}
\newcommand{\LRt}{\mathrm{\mathbf{T}_{LR}}}
\newcommand{\HRhat}{\mathrm{\mathbf{\hat{I}}_{HR}}}
\newcommand{\HRthat}{\mathrm{\hat{\mathbf{T}}_{HR}}}
\newcommand{\Net}{\mathcal{N}}
\newcommand{\T}{\mathcal{T}}
\newcommand{\Tinv}{\mathcal{T}^{-1}}
\newcommand{\Tq}{\mathcal{T}_\text{E}}
\newcommand{\Ts}{\mathcal{T}_\text{S}}

\newacronym{voc}{VOC}{Volatile Organic Compound}
\newacronym{bvoc}{BVOC}{Biogenic Volatile Organic Compound}
\newacronym{cnn}{CNN}{Convolutional Neural Network}
\newacronym{sr}{SR}{Super-Resolution}
\newacronym{sisr}{SISR}{Single-Image Super-Resolution}
\newacronym{san}{SAN}{Second-order Attention Network}
\newacronym{nlrg}{NLRG}{Non-Locally Enhanced Residual Group}
\newacronym{lsrag}{LSRAG}{Local-Source Residual Attention Groups}
\newacronym{srgan}{SRGAN}{Super-Resolution Generative Adversarial Network}
\newacronym{esrgan}{ESRGAN}{Enhanced SRGAN}
\newacronym{srresnet}{SRResNet}{Super-Resolution ResNet}
\newacronym{msrresnet}{MSRResNet}{Modified SRResNet}
\newacronym{rcan}{RCAN}{Residual Channel Attention Network}
\newacronym{rrdbnet}{RRDBNet}{Residual-in-Residual Dense Block Network}
\newacronym{sansisr}{SANSISR}{Second-order Channel Attention Network}
\newacronym{srcnn}{SRCNN}{SR Convolutional Neural Network}
\newacronym{cdf}{CDF}{Cumulative Distribution Function}
\newacronym{pdf}{PDF}{Probability Density Function}
\newacronym{hr}{HR}{High Resolution}
\newacronym{lr}{LR}{Low Resolution}
\newacronym{megan}{MEGAN}{Model of Emissions of Gases and Aerosols from Nature}
\newacronym{mse}{MSE}{Mean Squared Error}
\newacronym{nmse}{NMSE}{Normalized Mean Squared Error}
\newacronym{ssim}{SSIM}{Structural Similarity Index Measure}
\newacronym{dl}{DL}{Deep Learning}
\newacronym{lst}{LST}{Land Surface Temperature}
\newacronym{sst}{SST}{Sea Surface Temperature}
\newacronym{vgg}{VGG}{Visual Geometry Group}
\newacronym{pinn}{PINN}{Physics-Informed Neural Networks}
\newacronym{ctm}{CTM}{Chemistry Transport Model}

\title{Super-Resolution of BVOC Maps by Adapting Deep Learning Methods}
\name{Antonio Giganti, Sara Mandelli, Paolo Bestagini, Marco Marcon, Stefano Tubaro
\thanks{This work was supported by the Italian Ministry of University and
Research (MUR) and the European Union (EU) under the PON/REACT project.}}
\address{Dipartimento di Elettronica, Informazione e Bioingegneria - Politecnico di Milano - Milan, Italy}

\begin{document}
\ninept
\maketitle
\begin{abstract}
\glspl{bvoc} play a critical role in biosphere-atmosphere interactions, being a key factor in the physical and chemical properties of the atmosphere and climate.
Acquiring large and fine-grained \gls{bvoc} emission maps is expensive and time-consuming, so most available \gls{bvoc} data are obtained on a loose and sparse sampling grid or on small regions. However, high-resolution \gls{bvoc} data are desirable in many applications, such as air quality, atmospheric chemistry, and climate monitoring.
In this work, we investigate the possibility of enhancing \gls{bvoc} acquisitions, further explaining the relationships between the environment and these compounds.
We do so by comparing the performances of several state-of-the-art neural networks proposed for image \gls{sr}, 
adapting them to 
overcome the challenges posed by the large dynamic range of the emission and reduce the impact of outliers in the prediction. Moreover, we also consider realistic scenarios, considering both temporal and geographical constraints. Finally, we present possible future developments regarding \gls{sr} generalization, considering the scale-invariance property and super-resolving emissions from unseen compounds.  
\end{abstract}

\glsresetall


\section{Introduction}
\label{sec:intro}
Terrestrial ecosystems generate many chemicals, including volatile and semi-volatile compounds released into the atmosphere. Some of them, such as \glspl{bvoc}, play critical roles in atmospheric chemistry~\cite{cai_bvoc_scientometric_2021, guenther_model_2012}.
Their oxidation in the atmosphere affects tropospheric photochemistry and composition~\cite{sindelarova_high-resolution_2022}. \gls{bvoc}'s oxidation products promote the formation of low-level ozone and secondary organic aerosols, thus significantly impacting air quality and the Earth's radiative budget~\cite{ciccioli_impact_2023, opacka_isoprene_2021}.

Quantitative estimations of \gls{bvoc} emissions are required for numerical evaluations of past, current, and future air quality and climate conditions~\cite{cai_bvoc_scientometric_2021, guenther_model_2012, hewitt_bvoc_quantification_2011}. \gls{bvoc} emissions are routinely included in coupled climate and chemistry models such as regional and global air quality and Earth system models~\cite{cmaq_model}.
Different ground-based measurement techniques can be applied to acquire \gls{bvoc} emissions at diverse scales, from leaf to regional and global levels~\cite{opacka_isoprene_2021, hewitt_bvoc_quantification_2011}. However, available measurements are
limited in space and time; therefore, they might not be fully suitable to 
perform reliable simulations of atmospheric, climate, and forecasting models. 

We propose to overcome this issue by generating a denser spatial grid of \gls{bvoc} emissions, starting from a coarser one. We formulate our goal as an image \gls{sr} task, a general problem of computationally enhancing the resolution of a digital image. 
Due to the ill-posed nature of this task,
the goal of \gls{sr} is to 
constrain the problem of 
finding a unique mapping between a \gls{lr} image and its \gls{hr} counterpart in such a way that the \gls{lr} image is upscaled with high fidelity. 

Several \gls{sr} approaches have been proposed in the literature. Before the \gls{dl} era, ``classical'' methods were based on different working principles (e.g., sparse neighbor embedding, 
edge sharpening,
 etc.). Starting with the seminal work of \gls{srcnn}~\cite{dong_srcnn_2014}, \gls{sr} methods based on \gls{dl} are nowadays widely exploited, proving superior to classical methods. 


Applying image \gls{sr} techniques to \gls{bvoc} emissions is not straightforward. Indeed, many works focus on enhancing images that are evaluated by visual inspection (like photographs and biomedical data). However, \gls{bvoc} data have a diverse nature, being physical measures linked to a meaningful measurement unit (e.g., we cannot observe negative emissions). Moreover, \gls{bvoc} dynamic range is strongly different compared to classic 8-bit imagery. 
All these specific characteristics make the adaption of standard \gls{sr} methods a necessary operation for super-resolving \gls{bvoc} emissions. 

Dealing with data different from natural images is a well-known issue for researchers who apply \gls{sr} algorithms to enhance the quality
of 
satellite data~\cite{salvetti_misr_2020}. 
For example, authors in \cite{nguyen_sisr_lst_2022} focused on improving the resolution of land surface temperature,
while, recently, \gls{sr} of sea surface temperature has been addressed in~\cite{ping_sst_2021, izumi_sisr_sst_2022, lloyd_misr_optically_2022}. 
The improvement of trajectory calculations of wind fields 
has been tackled in~\cite{brecht_sisr_wind_2022}, where authors implemented an 
attention mechanism to increase \gls{sr} performances. 
Attention has also been used in~\cite{yasuda_micrometeorology_2022} to face the \gls{sr} of near-surface temperature. 
In~\cite{liu_air_pollution_2019}, the authors exploited external factors and spatiotemporal dependencies of a pollution field to increase its resolution, and a similar problem was addressed
in~\cite{bessagnet_sr_ctm_2021}.

In this work, we tackle the problem of \gls{sr} of \gls{bvoc} emission maps. 
To the best of our knowledge, no prior studies are facing this task.
The proposed investigations could provide dense data 
for the atmospheric chemical, climate, and air quality models. 
In addition, upsampling biogenic emission maps might be helpful for a variety of tasks, e.g., to capture small-scale processes, to improve characterization of the complex interaction between \glspl{bvoc} and other chemical compounds, and to better quantify  emissions induced from abiotic~\cite{akira_exchanges_bvoc_2021, feldner_abiotic_2022}
 and ozone stress.

To solve our goal, we explore the potential of \gls{dl},
investigating several state-of-the-art neural networks for \gls{sr}.
In particular, we study how to adapt these methods to the physical properties and the extremely wide dynamic range typical of \gls{bvoc} emissions, investigating specific data transformations. 
The proposed method enables the synthesis of emission maps, making a significant step toward fulfilling the wishes of atmospheric chemical and climate modeling communities.
Our main contributions are as follows:
\begin{itemize}[leftmargin=*]
\vspace{-4pt}
    \setlength\itemsep{0.1pt}
    \item We explore seven \gls{dl}-based \gls{sr} algorithms to enhance the resolution of \gls{bvoc} maps by a scale factor up to $4$;
    \item We propose an appropriate data transformation to adapt \gls{sr} methods to the \gls{bvoc} domain by exploiting statistical information extracted from emissions;
    \item We work with an extremely recent emission inventory, including highly-resolved biogenic global emissions~\cite{sindelarova_high-resolution_2022}; 
    \item We investigate the generalization capabilities of the proposed methodology, involving spatial and temporal constraints, different spatial resolutions and compounds. 
\end{itemize}
\vspace{-4pt}


\section{BVOC Super Resolution}
\label{sec:proposed method}
\glsreset{hr}
\glsreset{lr}
\glsreset{sr}

\begin{figure}[t]
\centering
    \begin{subfigure}[b]{0.31\textwidth}
        \centering
        \includegraphics[width=\textwidth]{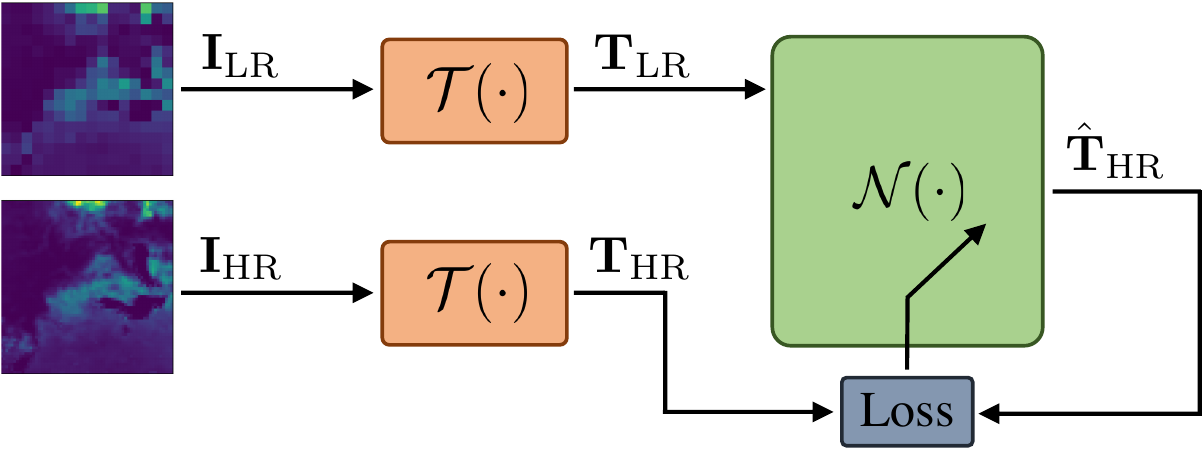}
        \caption{Training.}
        \label{fig:training}
    \end{subfigure}
\hfill
    \begin{subfigure}[b]{\columnwidth}
        \centering
        \includegraphics[width=\textwidth]{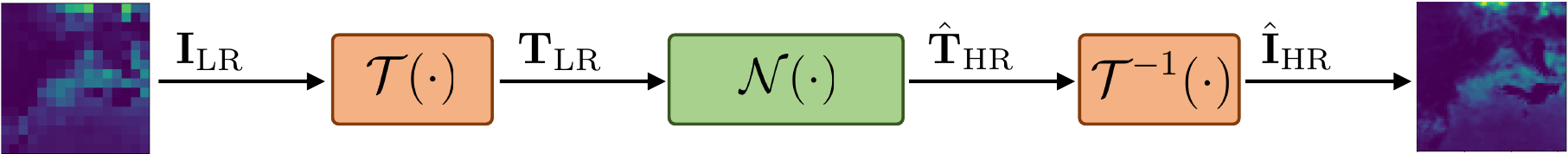}
        \caption{Deployment.}
        \label{fig:deployment}
    \end{subfigure}
\caption{The proposed method: (a) training and (b) deployment phase.}   
\label{fig:sisr_system}
  \vspace{-12pt}
\end{figure}

In this work, we consider the problem of recovering a \gls{hr} \gls{bvoc} emission map $\HR$ from a \gls{lr} emission map $\LR$, which can be seen as an image \gls{sr} task.
Fig.~\ref{fig:sisr_system} shows 
a sketch
of the proposed methodology. We estimate a super-resolved emission $\HRhat$ starting from the \gls{lr} emission $\LR$ as
\begin{equation}
    \HRhat = \Tinv(\Net(\T(\LR))),
\end{equation}
where $\T(\cdot)$ is a transformation applied to the \gls{lr} emission, $\Net(\cdot)$ is the operator applied by a neural network, 
and $\Tinv(\cdot)$ is the inverse transformation of $\T(\cdot)$.
We model $\LR$ as a matrix with size $M\times N$, while $\HR$ and its estimation $\HRhat$ have size $\alpha M\times \alpha N$, with $\alpha > 1$ indicating the \gls{sr} factor (i.e., how much we increase the resolution). 

The network training phase involves ($\LR$, $\HR$) emission pairs as input (see Fig.~\ref{fig:training}).
We consider the \gls{mse} between the transformed emissions as a loss function.
At testing stage (see Fig.~\ref{fig:deployment}), given an $\LR$ emission to be super-resolved, we estimate $\HRthat$ from the network. 
Then, we return to the original emission's range by applying the inverse data transformation $\Tinv(\cdot)$, 
obtaining the super-resolved emission $\HRhat$.

\vspace{-8pt}
\subsection{Data Transformation}
\label{ssec:data_transformation}

\gls{bvoc} emissions are spatially sparse, reporting extremely small values and a wide dynamic range (from $10^{-30}$ to $10^{-9}$ [kg/m\textsuperscript{2}s]~\cite{sindelarova_high-resolution_2022}).
Transforming this dynamic into more feasible values is crucial 
for numerical stability when training the network. 
To obtain emission maps with a dynamic range in $[0, 1]$, we 
investigate
two data transformation strategies, obtaining $\LRt = \TLR$ and $\HRt = \THR$, respectively:
\begin{itemize}[leftmargin=*]
\vspace{-3pt}
    \setlength\itemsep{0.1pt}
    \item $\Ts$ — \emph{Emission Scaling}: the transformed emissions are obtained by dividing $\LR$ and $\HR$ by their maximum, i.e., 
    \begin{equation}
         \LRt = \LR / \LR_{\max}, \quad \HRt = \HR / \HR_{\max};   
         \label{eq:scaling}
    \end{equation}


    \item $\Tq$ — \emph{Emission Equalization}: $\LRt$ and $\HRt$ are obtained by transforming $\LR$ and $\HR$ values to follow a uniform distribution between $0$ and $1$~\cite{scikit-learn}. In particular, we propose to learn the data transformation at training stage from \gls{hr} emission maps.  
    In a nutshell, we extract information on the \gls{cdf} of training data by subdividing them into different quantiles, associating a specific \gls{cdf} value with each training data quantile. Then, we apply this \gls{cdf} to all the emission maps for transforming them.
    To be more precise, for every possible emission value $K$ of \gls{hr} maps belonging to the training set, we can define 
    the associated \gls{cdf} value as
    \begin{equation}
        F(K) = \sum_{k=K_{\min}}^{K} p_k,
        \label{eq:cdf}
    \end{equation}
    where $K_{\min}$ is the lowest observed \gls{hr} emission value during training and $p_k$ is the probability that training \gls{hr} maps assume the value $k$. 
    We compute the transformed emission maps $\HRt$ and $\LRt$ as
    \begin{equation}
       [\HRt]_{i,j} = F([\HR]_{i,j}), \quad [\LRt]_{i,j} = F([\LR]_{i,j}), 
       \label{eq:quantile_transformer}
    \end{equation}
    where $i,j$ are the row and column coordinates of the emission values in each map. Notice that, since $F(K) \in [0, 1] \, \forall K$, the transformed emission maps also cover this value range. 
 \end{itemize}

Most of the past works in \gls{sr} for 2D data that involve physical quantities
simply scale the input according to the $\Ts$ approach~\cite{nguyen_sisr_lst_2022, izumi_sisr_sst_2022, yasuda_micrometeorology_2022}, 
discarding the intrinsic correlation of adjacent values. 
On the contrary, 
it has been proven that 
transforming data using statistical information extracted from quantiles of the input data distribution (i.e., $\Tq$ transformation) is less influenced by outliers than $\Ts$ preprocessing~\cite{scikit-learn, bogner_nqt_2012}
and can lead to better predictions
~\cite{peterson_oqn_2020}. 
This property is beneficial for \gls{bvoc} emissions since there can be many outliers due to
the large spatial diversity of the environmental factors driving the emission process, such as meteorology, type of vegetation, seasonal cycle, and atmospheric composition~\cite{sindelarova_high-resolution_2022}. 

Another difference between the two transformations is related to retrieving information on the maximum peak emission.
Indeed, due to the inherent low resolution of \gls{lr} maps, it is common to miss out the maximum emission value of \gls{hr} maps in their related \gls{lr} version.
In other words, it may happen very often that $\LR_{\max} < \HR_{\max}$, for each $(\LR, \HR)$ pair.
If this happens, $\Ts$ provides a processed version of $\LR$ not consistent with the data distribution in $\HR$.
On the contrary, $\Tq$ performs a non-linear transformation based on \emph{a priori} information derived from statistical analysis of the \gls{hr} data.
This is not sensitive to single outliers or local maxima in the data, making the distribution of $\LR$ and $\HR$ transformed data compatible.



\vspace{-8pt}
\subsection{SR Networks}
We compare seven different state-of-the-art neural networks for image \gls{sr}, investigating the adoption of different configurations and learning paradigms. 
The investigated networks are the following:
\begin{itemize}[leftmargin=*]
    \setlength\itemsep{0.1pt}
    \item \gls{srcnn}~\cite{dong_srcnn_2014}, which is a pioneer work based on \glspl{cnn};
    \item \gls{srgan}~\cite{ledig_srgan_2017}, where the authors define a novel perceptual loss based on feature maps of the \gls{vgg} network;
    \item \gls{rcan}~\cite{zhang_rcan_2018}, which is an attention-based framework. The recursive residual design allows multiple pathways for information flow from initial to final layers, and selective attention allows focusing on specific feature maps that are more important for the end task;
    \item \gls{esrgan}~\cite{wang_esrgan_2019}, which is an improved version of \gls{srgan}~\cite{ledig_srgan_2017}, including a different architecture for both the generator and the discriminator;
    \item \gls{rrdbnet}~\cite{wang_esrgan_2019}, combining 
    multi-level residual networks and dense connections; 
    \item \gls{sansisr}~\cite{dai_sansisr_2019}, where \gls{san} is proposed for more powerful 
    feature correlation learning, using second-order feature statistics. 
    The authors propose 
    an additional block to capture long-distance spatial information
    and to exploit \gls{lr} information, easing the training and bypassing low-frequency characteristics;
    \item \gls{msrresnet}~\cite{zhang_aim_2019}, which is a modified version of the original \gls{srresnet}~\cite{ledig_srgan_2017}, exploiting the benefits of the residual learning framework.
    
\end{itemize}
\section{Experimental Setup}
\label{sec:setup}

\subsection{Dataset}
\label{ssec:dataset}
Inventories of real measurements of \gls{bvoc} emissions are scarce and limited in time and space. 
However, the knowledge obtained from observations on the emission processes gives the possibility to simulate them for a specific temporal and spatial domain based on defined input parameters.
We use the most recent global coverage biogenic emission inventory calculated by an emission model and presented in~\cite{sindelarova_high-resolution_2022}. 
The inventory includes emissions from several biogenic compounds, covering the entire Earth surface from 2000 to 2019, with a $0.25^{\circ}\times0.25^{\circ}$ spatial resolution ($\approx28$km$\times28$km of the planet surface for each cell in continental regions). As far as we know, this is the most up-to-date global coverage biogenic emission inventory with the highest spatial resolution available in the literature. Emissions are reported as hour profiles and are averaged monthly.

Among the different biogenic compounds 
in the inventory, we select isoprene since it is by far the most important in terms of both global emission and atmospheric impact~\cite{opacka_isoprene_2021, sindelarova_high-resolution_2022}, with an annual emission of about half of the total \glspl{bvoc} emissions~\cite{akira_exchanges_bvoc_2021}.

Each emission map has 
a grid of $1440\times720$ cells. 
We slice them to obtain smaller non-overlapped patches of size $64\times64$ cells. This makes the training more computationally manageable and allows us to assume that there are not many radial distortions due to the Earth curvature in the area. Since these maps are spatially sparse, 
a considerable amount contains zero emissions. 
Thus, we discard the patches with non-zero emissions below $5\%$, avoiding unnecessary computation on zero-emission areas.
We end up with $81957$ different \gls{hr} patches $\HR$ that can be considered as ground truth.
We generate the associated \gls{lr} patches by performing bicubic downsampling, obtaining the $\LR$ emission maps of $16\times16$ cells.
The ($\LR$, $\HR$) pairs constitute our final dataset $\mathcal{D} = \{\HR_i, \LR_i\}$, for $i = 1, \dots, 81957$.
We aim at super-resolving \gls{lr} emission maps, which have a $1^{\circ}\times1^{\circ}$ spatial resolution, into \gls{hr} emission maps with $0.25^{\circ}\times0.25^{\circ}$ resolution, thus with scale factor $\alpha=4$.

\vspace{-8pt}
\subsection{Training Setting}
\label{ssec:training}

As reported in Sec.~\ref{ssec:data_transformation}, $\Tq$ transformation requires the estimation of the \gls{cdf} of \gls{hr} training data for mapping the emission probability distribution to the uniform one~\cite{scikit-learn}. We estimate this \gls{cdf} by exploiting all \gls{hr} training data, dividing their dynamics into $1000$ quantiles. Then, we preprocess $\HR$ and $\LR$ emissions with this estimated transformation. We keep this transformation fixed for what concerns further network training and deployment stages.

To train the networks, we divide our dataset into train, validation, and test sets with $70/20/10$ percentage amount, respectively.
We combine the ADAM optimizer with the Cosine Annealing (with restarts) learning rate scheduler, which enables to obtain faster computations and better results~\cite{loshchilov_decoupled_2019}.
We use $50k$ iterations with restarts equal to $1$ at iteration $10k, 20k$, and $40k$. The initial learning rate is $10^{-4}$, 
with a 
minimum value of $10^{-7}$. 
\section{Results}
\label{sec:results}

\subsection{Metrics}
\label{ssec:metrics}
To evaluate the performance of our method, we employ the \gls{ssim}, a well-known metric for quantifying the resemblance between two images. 
We also adopt the \gls{nmse}, defined as the \gls{mse} between $\HR$ and $\HRhat$, normalized by the average of $\HR^2$. 
A good result is the one with high \gls{ssim} and/or low \gls{nmse}.

\vspace{-8pt}
\subsection{Preliminary Studies}
\label{ssec:config selection}

\begin{figure}[t]
  \centering
  \includegraphics[width=0.9\columnwidth]{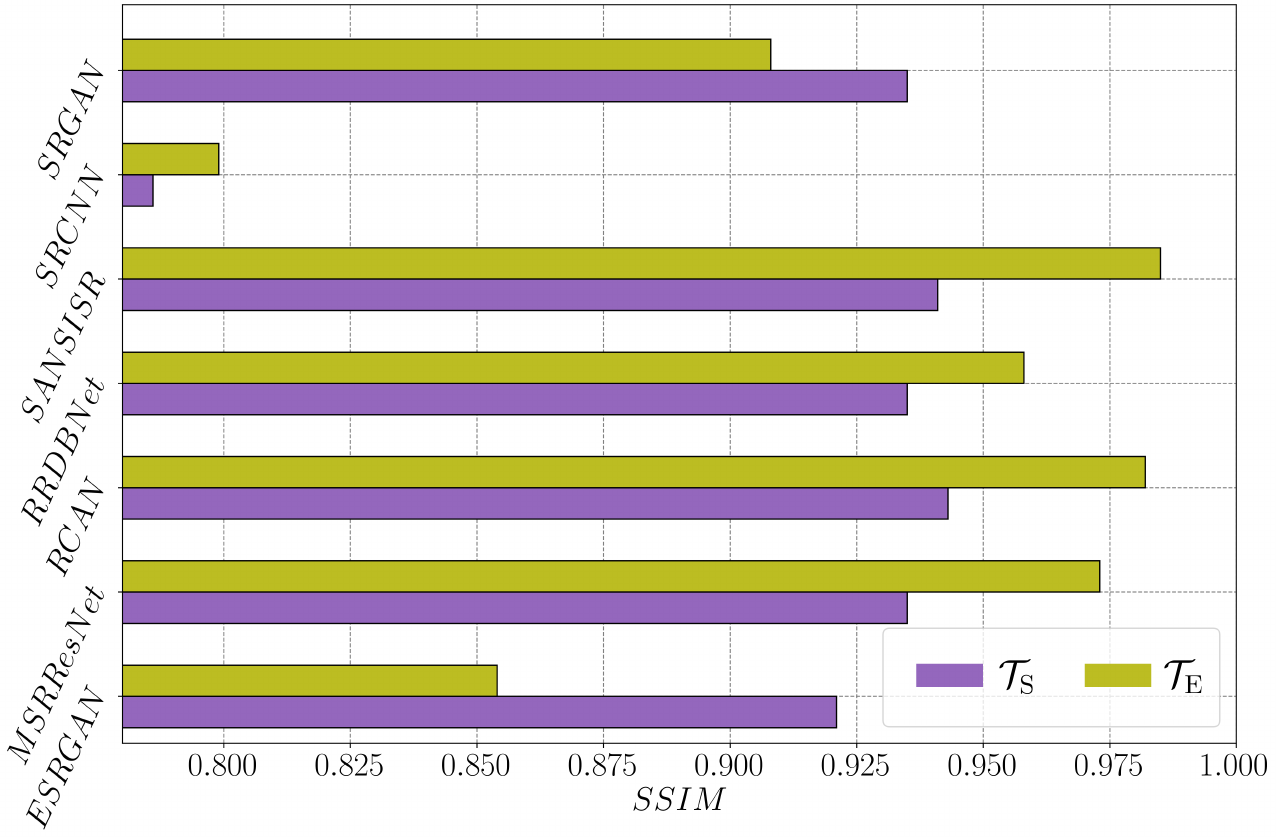}
  \caption{Average \gls{ssim} achieved by the different networks, for $\Ts$ and $\Tq$ data transformations.}
  \label{fig:param_selection}
  \vspace{-12pt}
\end{figure}

\begin{figure*}[t]
  \centering
  \includegraphics[width=.95\textwidth]{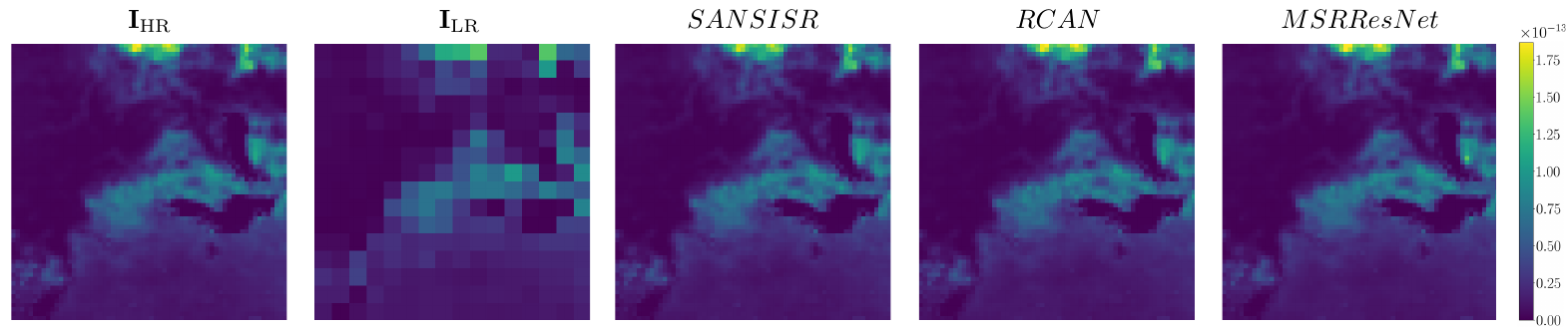}
  \caption{Image \gls{sr} examples of a generic \gls{bvoc} emission map from different algorithms. From left to right, the ground truth \gls{hr} image, the input \gls{lr} image, and three super-resolved results.}
  \label{fig:sr_results}
    \vspace{-12pt}
\end{figure*}

We compare the networks' performance 
for a single scale factor $\alpha = 4$ and two different data transformations (i.e., $\Ts$ or $\Tq$).
Fig.~\ref{fig:param_selection} summarizes the results in terms of \gls{ssim}. We can notice a reasonably high quality on all the architectures except for the \gls{srcnn}, with a significant boost on attention-based networks (i.e., \gls{sansisr} and \gls{rcan}).
It is worth noticing that $\Tq$ preprocessing almost always reports better reconstruction results than $\Ts$ strategy.
Given these reasons, 
we adopt $\Tq$ transformation 
in all the remaining experiments. 
In Fig.~\ref{fig:sr_results}, we report an example of super-resolved emission maps from the best-performing networks. 

\vspace{-8pt}
\subsection{Solving Realistic Scenarios}
\label{ssec:realistic_scenario}
In this section, we
investigate the performance of our method over more realistic and challenging scenarios. 
In particular, we aim to perform \gls{sr} of emission profiles in a different time range and geographical areas than those used in the training phase.
In a real-world context, the possibility to train with past data and test on future data would be a highly desirable feature.
Moreover, the availability of training data inherent to the same geographic area may be infeasible. As mentioned previously, measurements of biogenic emissions are often limited to restricted geographical areas.
To address these scenarios, from the main dataset $\mathcal{D}$ we build two subsets:
\begin{itemize}[leftmargin=*]
\vspace{-3pt}
    \setlength\itemsep{0.1pt}
    \item $\mathcal{D}_T$ — \emph{Time}:  we consider patches from years 2000-2014 for training, from years 2014-2018 for validation and years 2018-2019 for testing. This prevents the network from learning inter-annual biases and enables to investigate the temporal dependence.
    \item $\mathcal{D}_{TA}$ — \emph{Time \& Area}: a split version of $\mathcal{D}_T$ dataset that considers a portion of the spatial coverage for training and the remaining part for testing. This enables to investigate spatial dependence.
\end{itemize}
We compare the performances of our method over $\mathcal{D}_T$ and $\mathcal{D}_{TA}$ with those achieved on the initial dataset $\mathcal{D}$.
Notice that $\mathcal{D}_{TA}$ contains the smallest amount of emission pairs due to temporal and geographical constraints. 
To fairly compare the three scenarios, we randomly sample the bigger datasets $\mathcal{D}_T$ and $\mathcal{D}$ to have the same cardinality as $\mathcal{D}_{TA}$.
For simplicity, we compare only the three best-performing networks namely, \gls{sansisr}, \gls{rcan}, and \gls{msrresnet}.

\begin{table}[t]
\caption{Performance results over the datasets $\mathcal{D}$, $\mathcal{D}_T$ and $\mathcal{D}_{TA}$.}
\label{tab:realistic scenario}
\centering
\resizebox{\columnwidth}{!}{
\begin{tabular}{@{}lllll@{}}
\toprule
                   &  & \multicolumn{1}{c}{SANSISR}   & \multicolumn{1}{c}{RCAN}      & \multicolumn{1}{c}{MSRResNet} \\ \cmidrule(l){3-5} 
Dataset             &  & \multicolumn{1}{c}{SSIM / NMSE [dB]} & \multicolumn{1}{c}{SSIM / NMSE [dB]} & \multicolumn{1}{c}{SSIM / NMSE [dB]} \\ \midrule
$\mathcal{D}$    &  & $0.984$ / $-19.35$                  & $0.982$ / $-19.26$                  & $0.970$ / $-17.27$                  \\
$\mathcal{D}_T$    &  & $0.986$ / $-21.50$                  & $0.985$ / $-20.44$                  & $0.973$ / $-17.66$                  \\
$\mathcal{D}_{TA}$ &  & $0.769$ / $-7.31$                   & $0.774$ / $-7.32$                   & $0.788$ / $-7.26$                   \\ \bottomrule
\end{tabular}

}
  \vspace{-12pt}
\end{table}

Results are shown in Table~\ref{tab:realistic scenario}. The three networks perform very similarly, with \gls{sansisr} achieving the best results in most cases.
It is worth noticing that we can super-resolve unseen emissions from different years (i.e., $\mathcal{D}_T$ results). 
On the contrary, 
the performance drops drastically when dealing with geographical areas unseen in the training phase (i.e., $\mathcal{D}_{TA}$ results). This worse result occurs because of the close link between biogenic emissions and land morphology and the vegetation type of the area used in training. 

\vspace{-8pt}
\subsection{Towards Generalization in Emissions Super-Resolution}
\label{ssec:towards generalization}

In this section, we address two relevant problems related to the application of 
\gls{sr} methods on \gls{bvoc} emission maps.
In particular, we investigate 
the scale-invariance and the estimation of different compounds. We only show results related to \gls{sansisr} architecture, being it the best-performing one among the employed networks.

\textbf{Scale-invariance.} This property concerns the ability to generalize \gls{sr} to data with a spatial resolution not present in training. This could be a useful feature of \gls{sr} algorithms embedded in the climate modeling frameworks.
For instance, if we train the network to pass from $1.00^{\circ}\times1.00^{\circ}$ to $0.25^{\circ}\times0.25^{\circ}$ spatial resolution (i.e., with a $\alpha=4$ scale factor), we aim at testing the network capabilities in super-resolving emissions from $2.00^{\circ}\times2.00^{\circ}$ to $0.50^{\circ}\times0.50^{\circ}$ (i.e., the scale factor remains the same but the spatial resolution changes).

We investigate 
the scale-invariance 
in super-resolving emissions at two scale factors, i.e., $\alpha=2$ and $\alpha=4$.
Table~\ref{tab:generalization}(a) reports the experimental results.
As expected, we achieved excellent performance testing data with the exact spatial resolution used in training.
Whereas, 
for both $\alpha=2$ and $\alpha=4$ scale factors, testing over unseen resolutions leads to significant drops in performance.
This indicates that scale-invariance hypotheses are not fully satisfied.

\begin{table}[t]
\caption{Results for the scale-invariance (a) and for the estimation of different compounds (b). The configuration used during the training phase is reported in bold.}
\label{tab:generalization}
\resizebox{\columnwidth}{!}{
\subfloat[]{
\begin{tabular}{@{}lcl@{}}
\toprule
\begin{tabular}[c]{@{}l@{}}Scale \\ Factor\end{tabular} & \multicolumn{1}{c}{\begin{tabular}[c]{@{}c@{}}Spatial \\ Resolution\end{tabular}} & SSIM / NMSE [dB]    \\ \midrule
\multirow{4}{*}{$\alpha=2$}  & $\mathbf{0.50^{\circ}\rightarrow 0.25^{\circ}}$ & $0.993$ / $-25.58$ \\
 & & \\
& $1.00^{\circ}\rightarrow 0.50^{\circ}$ & $0.834$ / $-8.41$  \\
& $2.00^{\circ}\rightarrow1.00^{\circ}$ & $0.659$ / $-5.24$  \\ \midrule
\multirow{3}{*}{$\alpha=4$} 
& $\mathbf{1.00^{\circ}\rightarrow 0.25^{\circ}}$  & $0.985$ / $-20.86$ \\
& & \\
& $2.00^{\circ}\rightarrow 0.50^{\circ}$ & $0.603$ / $-4.84$  \\ \bottomrule
\end{tabular}
}
\quad
\subfloat[]{
\begin{tabular}{@{}lll@{}}
\toprule
\begin{tabular}[c]{@{}l@{}}Biogenic \\ Species\end{tabular} &  & SSIM / NMSE [dB]            \\ \midrule
$\bm{\mathcal{D}}$ &  & $0.985$ / $-20.87$         \\
&  & \multicolumn{1}{l}{} \\
$\mathcal{D}_\text{mon}$                                         &  & $0.875$ / $-15.76$         \\
$\mathcal{D}_\text{met}$                                         &  & $0.752$ / $-13.30$         \\
$\mathcal{D}_\text{ses}$                                         &  & $0.905$ / $-16.03$         \\ \bottomrule
\end{tabular}
}
}
  \vspace{-12pt}
\end{table}

\textbf{Estimation of different compounds.} 
We investigate the 
\gls{sr} of emission maps utilizing compounds different than those adopted in training.
As the scale-invariance, this property could be a desirable feature
in atmospheric modeling frameworks for enhancing the resolution of a generic chemical compound.

We build three datasets similar to $\mathcal{D}$ but containing emissions of diverse compounds. 
We select monoterpenes ($\mathcal{D}_\text{mon}$), methanol ($\mathcal{D}_\text{met}$), and sesquiterpenes ($\mathcal{D}_\text{ses}$), since these are the most responsible  compounds for the majority of the global \gls{bvoc} emission~\cite{sindelarova_high-resolution_2022}.
We train our method over the reference dataset $\mathcal{D}$ and test on the three datasets.
Results are depicted in Table~\ref{tab:generalization}(b).
Given the similarity of their chemical structure with that of isoprene,
it is worth noticing that terpenoids like monoterpenes (i.e., $\mathcal{D}_\text{mon}$) and sesquiterpenes (i.e., $\mathcal{D}_\text{ses}$) could be used as input to our algorithm, with results that are still acceptable if we consider the benefits that \gls{sr} emission maps could give to air quality and climate models.

\section{Conclusions}
\label{sec:conclusion}
This work focused on enhancing the spatial resolution of \gls{bvoc} emission maps using \gls{dl} techniques, proposing a suitable data transformation that drastically reduces the impact of outliers and improve the accuracy and reliability of our models.
Super-resolution of \gls{bvoc} emission maps is critical since global inventories of
\gls{bvoc} measurements are currently unavailable to the modeling communities, and processing existing emissions appears to be the only way to produce maps with high spatial resolution.

The overall high performance of the proposed method
shows that we effectively
generalize to unseen data, achieving accurate estimations of potential future \gls{bvoc} emissions.
We also indicate possible avenues 
towards improvements in the algorithms' generalization, investigating scale-invariance, and testing different compounds.
In the future, we will extend our work through the use of \gls{pinn}, considering both data and the physics behind the process.
Super-resolved \gls{bvoc} emissions could represent an invaluable 
source for air quality and climate products, allowing 
us to explain better how these compounds affect humans and global climate change.


\newpage

\section{References}
\printbibliography[heading=none]

\end{document}